\documentclass[aps,prc,reprint,amssymb,amsmath,superscriptaddress, floatfix,10pt]{revtex4-1}
\usepackage{mathptmx}
\usepackage{graphicx}
\usepackage{color}  
\makeindex
\newcommand{\textblue}{\textcolor[rgb]{0.00,0.07,1.00}}

\begin{document}

\title{The screening condition in the core of neutron stars}
\author{Dmitry Kobyakov}
\email{dmitry.kobyakov@appl.sci-nnov.ru}
\affiliation{Institute of Applied Physics of the Russian Academy of Sciences, 603950 Nizhny Novgorod, Russia}

\begin{abstract}
Earlier, the screening condition in neutron star core has been formulated as equality of velocities of superconducting protons and the electrons $\mathbf{v}_p=\mathbf{u}_e$ at wavenumbers $q\ll\lambda^{-1}$ ($\lambda$ is the London penetration depth) and has been used to derive the force exerted by the electrons on a moving flux tube.
By calculating the current-current response, I find that $\mathbf{v}_p\neq\mathbf{u}_e$ for $l^{-1}<q\ll\lambda^{-1}$ ($l$ is the electron mean free path).
I show that at typical realistic parameters the electric field induced by a moving (relative to the electrons) flux tube is not screened by the electron currents.
The implication is that the existing picture of the momentum exchange between the electrons and the flux tubes must be reassessed.
\end{abstract}

\maketitle

\emph{Introduction.}
In the outer core of neutron stars, one expects that the protons are superconducting, the electrons are normal and the magnetic field is present in the interior, which induces a magnetic flux tube lattice in the superconducting protons.
Scattering of the electrons by the flux tubes effectively couples the electrons and the protons, and the interaction of magnetized neutron vortices with the flux tubes effectively couples the superfluid neutrons and superconducting protons.
These couplings play an important role in modelling of observable phenomena in neutron stars.

Scattering of the electrons by a single flux tube (due to a neutron vortex) has been considered for the first time by Alpar, Langer and Sauls \cite{AlparEtal1984}.
In later works, the problem was generalized to the case of the electron scattering by the superconducting proton flux tubes and the flux tube interaction with the magnetized superfluid neutron vortices was considered \cite{RudermanZhuChen1998}.
Calculations were done with the hydrodynamic form of the electrical conductivity, see equation (10) in \cite{RudermanZhuChen1998}; the hydrodynamic form of the electrical conductivity was also used for the astrophysical applications in \cite{Jones1987,Jones1991}.
Later, controversial results have been obtained and attempts were made to resolve the controversy \cite{Gusakov2018} (and references therein).
In equation (16) of \cite{Gusakov2018}, the so called screening condition $\mathbf{v}_p=\mathbf{u}_e$ was a basic assumption in the derivations, where $\mathbf{v}_p$ is the velocity of the superconducting protons and $\mathbf{u}_e$ is the velocity of the electrons.
Notice that the derivations in \cite{Gusakov2018} are based on the assumption that the separation between the flux tubes $d_v$ is much larger than the size of the magnetic core $\lambda$ (figure 3 in \cite{Gusakov2018}).
however, straightforward calculations in the present paper show that in fact $d_v\sim\lambda$; this implies that in realistic conditions it is impossible to choose a surface for the integration of the momentum flux without a nonzero electric current (and, thus, $\mathbf{v}_p\neq\mathbf{u}_e$).
The screening condition was also used in \cite{SourieChamel2020} in studies of superfluid neutron vortices in the core of neutron stars.

The physical meaning of the screening condition is that sufficiently far from the flux tube the total electrical current is zero, $\mathbf{J}_{e}+\mathbf{J}_{p}=0$, where $\mathbf{J}_{e}$ is the electronic electric current and $\mathbf{J}_{p}$ is the supercurrent due to protons.
Applied to the realistic parameters in neutron stars, it means that the screening condition assumed in \cite{Gusakov2018,SourieChamel2020} must be satisfied in the spatial region between the flux tubes, where the superconductor density is spatially uniform.
In fact, this assumption is not obvious because the screening condition is expected to hold in the hydrodynamic regime, while at length scales of the order of separation distance between the flux tubes the electron regime is not hydrodynamic.

The hydrodynamic regime is realized when the relevant wavenumbers $q$ are much smaller than the electron inverse mean free path $l^{-1}$ \cite{EassonPethick1977}.
In fact, the electron mean free path $l$ is typically somewhere between $10^6$ fm and $10^{11}$ fm, as will be shown below.
For given magnetic field strength $H$, the typical separation distance between the flux tubes is
\begin{equation}\label{dv}
d_v=\left(\frac{\Phi_0}{H}\right)^{1/2}\approx 143.8\left(\frac{10^{15}\;\mathrm{Oe}}{H}\right)^{1/2}\;\mathrm{fm},
\end{equation}
where $\Phi_0=\pi\hbar c/e$ is the magnetic flux quantum, $c$ is the speed of light, $e$ is the proton electric charge.
At typical temperature $T=10^8$ K, the neutron star matter in the outer core is effectively in the zero temperature regime.
The length scale of the magnetic tube associated with the proton vortex is
\begin{equation}\label{lambda0}
\lambda(0)=\left(\frac{m_p c^2}{4\pi e^2 n_p}\right)^{1/2}\approx 80\;\mathrm{fm},
\end{equation}
where $n_p=x_pn$ is the proton number density, $n=0.16\;\mathrm{fm}^{-3}$ is the baryon number density, $x_p=0.05$ is the proton fraction.
The electron Fermi wavenumber is $k_e=(3\pi n_e)^{1/3}\approx0.6187\;\mathrm{fm}^{-1}$.

In this paper, I will show that when the flux tube size $\lambda$ is resolved by the electron dynamics (that is, at wavenumbers $q\sim\lambda^{-1}$), the electrons are in the particle-hole regime.
I will calculate the current-current response of the electrons and show that the electric fields arising as a result of the flux tube motion cannot be screened by the electrons leading to $\mathbf{J}_{e}+\mathbf{J}_{p}\neq0$ in the spatial region between the flux tubes, which implies that the screening condition is not satisfied.
This implies that the considerations, in particular, the magnetic field evolution \cite{RudermanZhuChen1998} based on the hydrodynamic form of the electron electrical conductivity should be reassessed.

\emph{Screening of the supercurrent by the electrons.}
I will do calculations with the material parameters corresponding to the nuclear saturation density.
The electron velocity $\mathbf{u}_e$ is the microscopic velocity averaged on length scale much longer than $k_e^{-1}$.
Likewise, the proton superflow velocity $\mathbf{v}_p$ is the quantity averaged on length scale much longer than the proton coherence length $\xi(T)$.
At $T\ll T_c$, the superconducting density is equal to $n_p$, where $T_c={\gamma\Delta_p}/{k_B\pi}$, with $\ln\gamma=0.577$ is the Euler constant and $\Delta_p$ is the proton s-wave superconducting energy gap, and
\begin{equation}\label{xi0D}
\xi(0)=\frac{\hbar v_{Fp}}{2\gamma\Delta_p}\approx7.2\left(\frac{1\,\mathrm{MeV}}{\Delta_p}\right)\;\mathrm{fm}.
\end{equation}
The coherence length $\xi$ sets the length scale for the normal core of the flux tube, while the magnetic core extends to the scale $\sim\lambda$.
Here, $v_{Fp}=\hbar k_p/m_p$ is the electron Fermi velocity, $k_p=k_e$ and $m_p$ is the proton mass.

The number currents are defined as $\mathbf{j}_{e}=n_{e}\mathbf{u}_e$ and $\mathbf{j}_{p}=n_{p}\mathbf{v}_p$ and the electric currents are $\mathbf{J}_{e}=-en_{e}\mathbf{u}_e$ and $\mathbf{J}_{p}=en_{p}\mathbf{v}_p$.
Noting that $m_e\ll m_p$, where $m_e$ is the electron mass, is a reasonable lowest-order approximation even for the relativistic electrons, {one} may neglect the inertia of the electrons \cite{GlampedakisEtal2011,Kobyakov2018}.

Therefore, for a given electric field strength, it is sufficient to assume that the proton velocity is given and then to calculate the equilibrium electron velocity.
In this context one is free to choose the proton phase as a constant, so the supercurrent is defined by the electromagnetic vector potential \cite{Kobyakov2018}.
It will be convenient to work in the Fourier representation, $\mathbf{X}(\mathbf{q},\omega)=\int d^3\mathbf{x}dt \,e^{i\mathbf{q}\cdot\mathbf{x}-i\omega t}\mathbf{X}(\mathbf{x},t)$, where $\mathbf{X}(\mathbf{x},t)$ is any function of space and time.
Thus, the proton current $\mathbf{J}_p$ may be seen as given which is denoted for convenience $\mathbf{J}_{\mathrm{ext}}$, and the electronic current $\mathbf{J}_e$ may be seen as the induced current $\mathbf{J}_{\mathrm{ind}}$; the total current is the sum $\mathbf{J}_p+\mathbf{J}_e$.
Since a given proton supercurrent is a transverse vector ($\nabla\cdot\mathbf{J}_{p}=0$) the electron response is defined by the transverse electromagnetic response of the system.

\emph{Description of given, induced and the total electric currents in terms of the dielectric function.}
Calculation of the electron current $\delta\mathbf{J}_{e}$ for a given proton current $\delta\mathbf{J}_{p}$ can be done in the lowest approximation, $\delta\mathbf{J}_{e}(\mathbf{q},\omega)=\sigma(\mathbf{q},\omega){\delta\mathbf{E}(\mathbf{q},\omega)}$.
The electric field ${\delta\mathbf{E}(\mathbf{q},\omega)}$ is related to the vector potential:
\begin{equation}\label{London1equation}
\mathbf{E}=-c\partial_t\mathbf{A}.
\end{equation}
Thus, to characterize the system it is necessary to calculate the quantity
\begin{equation}\label{Def1Sigma}
  \sigma(\mathbf{q},\omega)=\frac{\delta\mathbf{J}_{e}(\mathbf{q},\omega)}{\delta\mathbf{E}(\mathbf{q},\omega)}=\frac{ec}{\mathrm{i}\omega}\frac{\delta\mathbf{j}_{e}(\mathbf{q},\omega)}{\delta\mathbf{A}(\mathbf{q},\omega)},
\end{equation}
which is the electrical conductivity of the electrons.

The electrical conductivity may be expressed through the dielectric function $\epsilon_t$, for which one has at least two equivalent choices.
I will work with the transverse fields in this paper; for convenience, the subscript $t$ is added to stress that $\epsilon_t$ is the transverse dielectric function.
The definition of the dielectric function following Lindhard \cite{Linhard1954} (with the superscript ``L''), is based on a linear relation between the electric induction $\mathbf{D}$ and the electric field $\mathbf{E}$,
\begin{equation}\label{DefEpsilonLinhard}
\mathbf{D}=\epsilon_t^{(\mathrm{L})}\mathbf{E},
\end{equation}
and then the Maxwell equation can be written as
\begin{equation}\label{MaxwellEpsilonLindhard}
  \left( q^2-\frac{\omega^2}{c^2}\epsilon_t^{(\mathrm{L})}(q,\omega) \right)\mathbf{A}(q,\omega)=\frac{4\pi}{c}\mathbf{J}_{\mathrm{ext}}(q,\omega).
\end{equation}
Alternatively, the definition of the dielectric function following Jancovici \cite{Jancovici} is based on a linear relation between the total (self-consistent) vector potential $\mathbf{A}=\mathbf{A}_{\mathrm{ext}}+\mathbf{A}_{\mathrm{ind}}$ (where $\mathbf{A}_{\mathrm{ext}}$ is the external (given) part due to a source and $\mathbf{A}_{\mathrm{ind}}$ is the induced part due to the electrons) and the external vector potential:
\begin{equation}\label{DefEpsilonJancovici}
  \mathbf{A}(q,\omega)=\frac{1}{\epsilon_t(q,\omega)}\mathbf{A}_{\mathrm{ext}}(q,\omega).
\end{equation}
Then the Maxwell equation reads
\begin{equation}\label{MaxwellEpsilonJancovici}
  \left( q^2-\frac{\omega^2}{c^2}\right)\epsilon_t(q,\omega)\mathbf{A}(q,\omega)=\frac{4\pi}{c}\mathbf{J}_{\mathrm{ext}}(q,\omega).
\end{equation}
In terms of the identifications assumed above, the electronic current is a linear functional of the supercurrent:
\begin{equation}\label{DefJind}
  \mathbf{J}_{e}(\mathbf{q},\omega)=\frac{1-\epsilon_t(\mathbf{q},\omega)}{\epsilon_t(\mathbf{q},\omega)}\mathbf{J}_{p}(\mathbf{q},\omega).
\end{equation}
It is easy to relate the two definitions to one another:
\begin{equation}\label{epsilon_t_epsilon_t_L}
  \epsilon_t^{(\mathrm{L})}(\mathbf{q},\omega)-1=\frac{c^2q^2-\omega^2}{\omega^2}\left(1 - \epsilon_t(q,\omega)\right).
\end{equation}
It follows from the Maxwell equations that $\sigma$ is related to $\epsilon_t^{(\mathrm{L})}$ \cite{PinesNozieres}:
\begin{equation}\label{Def2Sigma}
  \sigma(\mathbf{q},\omega)=\frac{\omega}{\mathrm{i}4\pi}\left(\epsilon_t^{(\mathrm{L})}\left(\mathbf{q},\omega\right)-1\right).
\end{equation}
It is convenient to introduce the response function $\tilde{\chi}_t(\mathbf{q},\omega)$ according to the definition
\begin{equation}\label{DefChi}
\epsilon_t(\mathbf{q},\omega)=1+\frac{4\pi e^2}{q^2}\tilde{\chi}_t(\mathbf{q},\omega).
\end{equation}
Combining the above equations one can write
\begin{equation}\label{ChiDjDA}
\tilde{\chi}_t(\mathbf{q},\omega)=\frac{cq^2}{e(\omega^2-c^2q^2)}\frac{\delta \mathbf{j}_e(\mathbf{q},\omega)}{\delta\mathbf{A}(\mathbf{q},\omega)}.
\end{equation}

\emph{Calculation of }$\sigma(\mathbf{q},\omega)$.
The next step is to find the explicit form of $\epsilon_t(\mathbf{q},\omega)$.
In the present system, the electrons are relativistic and quantum degenerate.
It is important to distinguish between the regimes of the electron dynamics \cite{PinesNozieres}: the hydrodynamic regime is realized when $\omega\ll\nu$ and $q\ll l^{-1}$; the particle-hole regime results when $\omega>\nu$ and$/$or $q>l^{-1}$.
There are two options in the particle-hole regime: with $q>l^{-1}$ and $\omega>\nu$ the collisions are unimportant; with $q>l^{-1}$ and $\omega<\nu$ the collisions are important.
Here, $\nu=c/l$ is the electron collision rate, which is defined as a sum of the collision rates with all kinds of impurities that scatter the electrons.
In the present system,
\begin{equation}\label{DefNu}
  \nu=\nu_1+\nu_2,
\end{equation}
where $\nu_1$ is the rate of collisions of the electrons with the normal protons within the magnetic flux tubes ($\nu_2$ is the rate of collisions of the electrons with the magnetic field lines within the magnetic flux tubes and the superfluid neutron vortices).

I will focus on the short length scales ($q\gg l^{-1}$) and slow frequencies $\omega\ll\nu$.
Jancovici \cite{Jancovici}  has already calculated the dielectric function $\epsilon_t(\mathbf{q},\omega)$ for $q\geq l^{-1}$, but completely neglecting the electron collisions (for $\omega\gg\nu$).
At temperatures $T\ll T_c$ the normal protons in the bulk of the superconductor may be neglected and collisions of the electrons with the normal protons occur only within the flux tubes, therefore
\begin{equation}\label{Nu1}
\nu_1\approx\tau_{\mathrm{tr}}^{-1}\frac{H}{H_{c2}(0)},
\end{equation}
where $H_{c2}(0)={\Phi_0}/{2\pi\xi(0)^2}$, is the upper critical magnetic field of the superconductor at $T=0$, $H$ is the stellar magnetic field.
The electron transport relaxation time with normal protons was evaluated by Baym, Pethick and Pines \cite{BPP1969b} as $\tau_{\mathrm{tr}}\approx2\times10^{-14}\;\mathrm{s}$.
With typical $H=10^{15}$ Oe and $\Delta_p=1$ MeV, I find $\nu_1\approx7.898\times10^{11}\;\mathrm{s}^{-1}$.
For scattering of the electrons by the magnetic field in the flux tube, I use the order of magnitude estimate
\begin{equation}\label{Nu2}
\nu_2=\frac{c}{l_{\mathrm{et}}}\approx c\sigma_{\mathrm{et}}n_t,
\end{equation}
where $l_{\mathrm{et}}$ is the electron mean free path between consecutive collisions of the electron with the flux tube, $\sigma_{\mathrm{et}}$ is the differential cross-section for scattering of the electron with the flux tube and $n_t=H/\Phi_0$ is the number of flux tubes per unit area.
Note that if $H$ is smaller than the lower critical magnetic field, the neutron vortices would provide the dominant impurity scattering for the electrons.
The cross section is given by $\sigma_{\mathrm{et}}=\alpha k_e^{-1}$, where $\alpha=\alpha(k_e\xi,\lambda/\xi)$.
From equations (37) and (40) of \cite{Gusakov2018} I infer that $\alpha\sim10^{-2}$.
Thus, with typical $H=10^{15}$ Oe and $\Delta_p=1$ MeV, I find $\nu_2\approx\alpha\times2.343\times10^{19}\;\mathrm{s}^{-1}$ and $\nu\approx\nu_2$.
This coincides with the standard theoretical expectation that the main source of the electron-proton coupling in superconducting matter of neutron star core is the electron interaction with the magnetic flux tubes.
The quantitative estimates in Eqs. (\ref{Nu1}) and (\ref{Nu2}) enable to calculate the minimum ($c/\nu_2$) and the maximum ($c/\nu_1$) mean free path of the electrons, leading to $\sim10^6$ fm and $\sim10^{11}$ fm correspondingly, with typical $H=10^{15}$ Oe and $\Delta_p=1$ MeV.

In neutron stars, $\omega$ describes dynamical processes related to jumps in the rotational frequency of the star, or to mechanical torsional oscillations, or to sound waves and may take values in the range between $\sim0$ and $\sim10^4$ Hz.
Therefore $\omega\ll\nu$ for typical realistic conditions.

For the further calculations, $\epsilon_t(\mathbf{q},\omega)$ calculated by Jancovici \cite{Jancovici} must be modified in order to include the electron collisions.
I will use the following notation:
\begin{equation}\label{DefChiTilda}
  \tilde{\chi}_t(\mathbf{q},\omega)=\left\{\begin{array}{c}
                                           \chi_t(\mathbf{q},\omega) \;\mathrm{for}\;\omega\gg\nu,\\
                                           \chi_t^\nu(\mathbf{q},\omega) \;\mathrm{for}\;\omega\ll\nu.
                                         \end{array}
   \right.
\end{equation}

As the second step, I turn to the kinetic equation, from which the functional derivative
${\delta\mathbf{j}_{e}(\mathbf{q},\omega)}/{\delta\mathbf{A}(\mathbf{q},\omega)}$ can be calculated in both cases, when either $\omega\ll\nu$ (the collision integral $I[n_{1p}]$ is nonzero) or $\omega\gg\nu$ ($I[n_{1p}]=0$).
Here, $n_{1p}$ is the departure of the distribution function from true equilibrium \cite{PinesNozieres,ContiVignale1999}.
In the relaxation time approximation (RTA), the collision integral is written in the form
\begin{equation}\label{IRTA}
  I[n_{1p}]=-\nu\left(n_{1p}-n_{1p}^R\right),
\end{equation}
where $n_{1p}^R$ is the so-called locally relaxed equilibrium distribution function \cite{ContiVignale1999}.

As Conti and Vignale have shown \cite{ContiVignale1999}, in RTA the response function \emph{with} collisions ($\chi_t^\nu(\mathbf{q},\omega)$) can be obtained from the response function \emph{without} collisions with the frequency $\omega$ replaced by $\omega+\mathrm{i}\nu$ ($\chi_t(\mathbf{q},\omega+\mathrm{i}\nu)$):
\begin{equation}\label{ChiNu}
  \chi_t^\nu(\mathbf{q},\omega)=\frac{\omega}{\omega+\mathrm{i}\nu}\chi_t(\mathbf{q},\omega+\mathrm{i}\nu).
\end{equation}
Notice that Conti and Vignale \cite{ContiVignale1999} have worked with the quantity $\chi_t^\tau$ (the superscript 1999 is referring to the quantities used in \cite{ContiVignale1999}),
\begin{equation}\label{djdA1999}
\chi_t^\tau(\mathbf{q},\omega)\equiv\frac{\delta\mathbf{j}_{e}^{1999}(\mathbf{q},\omega)}{\delta\mathbf{A}^{1999}(\mathbf{q},\omega)}=-\frac{c}{e}\frac{\delta\mathbf{j}_{e}(\mathbf{q},\omega)}{\delta\mathbf{A}(\mathbf{q},\omega)},
\end{equation}
which, as can be easily seen from Eqs. (\ref{ChiDjDA}), (\ref{DefChiTilda}) and (\ref{djdA1999}), is proportional to $\chi_t^\nu$:
\begin{equation}\label{chiNuchiTauRelation}
  \chi_t^\nu(\mathbf{q},\omega)=\frac{q^2}{c^2q^2-\omega^2}\chi_t^\tau(\mathbf{q},\omega).
\end{equation}
By virtue of the proportionality between $\chi_t^\nu$ and $\chi_t^\tau$ seen in Eq. (\ref{chiNuchiTauRelation}), the result obtained in RTA for the relation between $\chi_t^\tau$ and its collisionless counterpart by Conti and Vignale \cite{ContiVignale1999}, is applicable to the relation between $\chi_t^\nu$ and $\chi_t$; this validates the formula in Eq. (\ref{ChiNu}).

This conclusion is useful because it allows to find $\chi_t^\nu(\mathbf{q},\omega)$ from $\chi_t(\mathbf{q},\omega)$, which has been obtained in \cite{Jancovici}.
Equation (66) from \cite{Jancovici} gives:
\begin{equation}\label{ChiJancovici}
  \chi_t(\mathbf{q},\omega)=\frac{s}{2}\frac{\partial n_e}{\partial \mu_e}\left(\frac{s}{1-s^2}+\frac{1}{2}\log\frac{s+1}{s-1}\right),
\end{equation}
where $\mu_e$ is the electron Fermi energy, ${\partial n_e}/{\partial \mu_e}=k_e^2/\pi^2\hbar c$ and $s\equiv s(q,\omega)=\frac{\omega}{cq}$.
Collecting the results I obtain the main formula of this paper:
\begin{equation}\label{SigmaResult}
  \sigma(\mathbf{q},\omega)=\frac{\omega^2-c^2q^2}{\mathrm{i}\omega}\frac{e^2}{q^2}\frac{\omega}{\omega+\mathrm{i}\nu}\chi_t(\mathbf{q},\omega+\mathrm{i}\nu).
\end{equation}
From Eqs. (\ref{DefChi}), (\ref{DefChiTilda}), (\ref{ChiNu}) and (\ref{ChiJancovici}) I will calculate the quantity
\begin{equation}\label{DefZeta}
\zeta\equiv\frac{1-\epsilon_t(\mathbf{q},\omega)}{\epsilon_t(\mathbf{q},\omega)},
\end{equation}
which indicates effectiveness of the current-current screening, see Eq. (\ref{DefJind}).
The limiting case, $\zeta=-1$ implies that $\mathbf{v}_p=\mathbf{u}_e$ -- the screening condition holds.
On the contrary, $\zeta=0$ implies that $\mathbf{v}_p\neq\mathbf{u}_e$ -- the screening condition is not satisfied.

\emph{Numerical results.}
\begin{figure}
\includegraphics[width=3.5in]{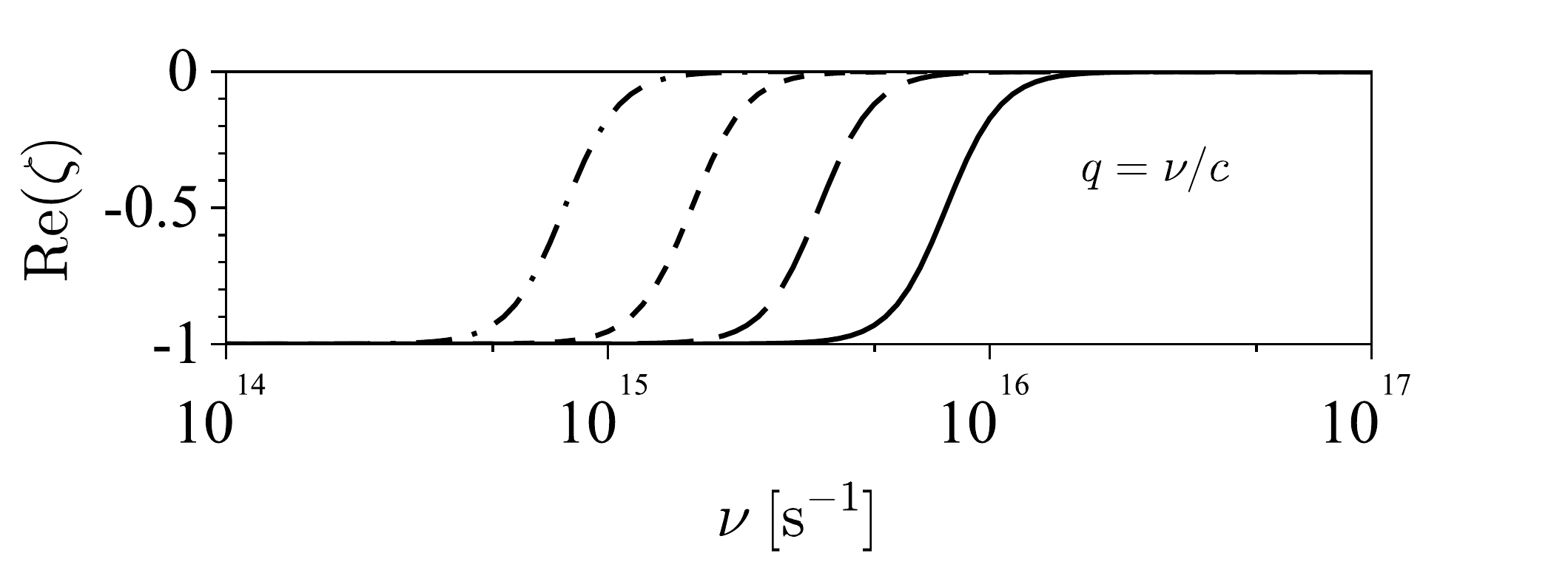}
\caption{The real part of the proportionality coefficient $\zeta$ between the induced current and the test current, Eq. (\ref{DefZeta}), as function of the collision frequency $\nu$ with wavenumber set as $q=\nu/c$. From left to right, the lines correspond to $\omega=10$, $10^2$, $10^3$ and $10^4$.}
\end{figure}
\begin{figure}
\includegraphics[width=3.5in]{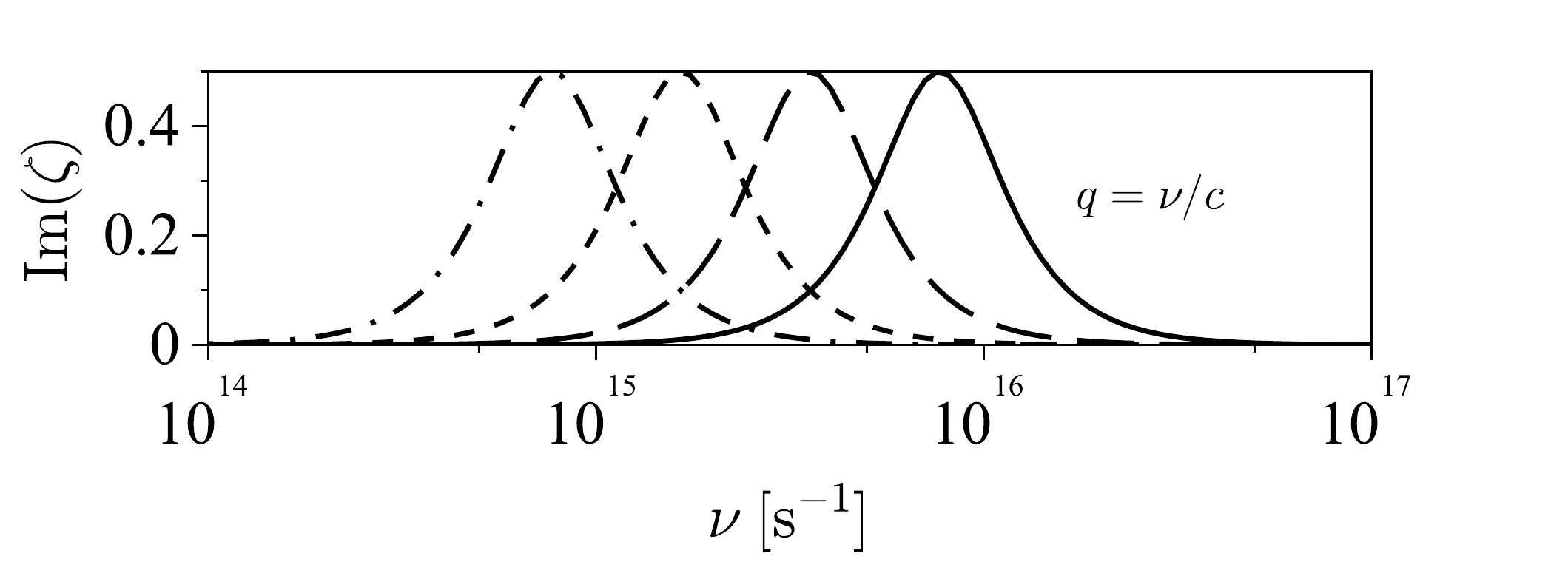}
\caption{The imaginary part of the proportionality coefficient $\zeta$ between the induced current and the test current, Eq. (\ref{DefZeta}), as function of the collision frequency $\nu$ with wavenumber set as $q=\nu/c$. From left to right, the lines correspond to $\omega=10$, $10^2$, $10^3$ and $10^4$.}
\end{figure}
I turn to evaluation of $\zeta$ for various $q$, $\omega$ and $\nu$.
Figures 1-4 display $\zeta$, Eq. (\ref{DefZeta}), as function of $\nu$ for four different frequencies ($\omega=10$, $10^2$, $10^3$ and $10^4$) and for two different length scales ($q=\nu/c$ and $10\nu/c$).
The chosen set is sufficient to reveal general patterns in behavior of $\zeta$ as function of $q$, $\omega$ and $\nu$.
Comparing the positions of lines in Figs. 1 and 3 {one observes} that at fixed $\nu$, decreasing of $q$ leads to improvement of screening; this implies that the smaller $q$ is, the closer $\zeta$ is to minus unity, as expected.
{One} can say that decreasing of $q$ at fixed $\omega$ moves the curve $\zeta$ in Fig. 1 to the right.
\begin{figure}
\includegraphics[width=3.5in]{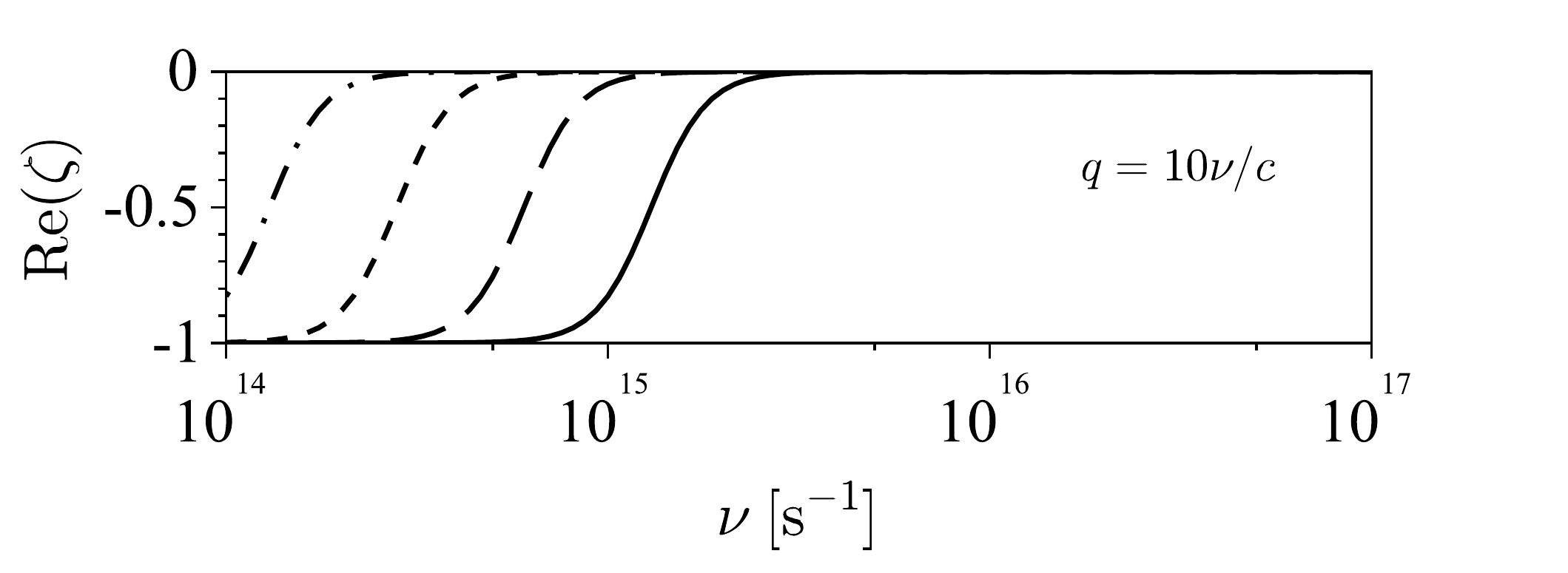}
\caption{The same as Fig. 1 but $q=10\nu/c$. }
\end{figure}
\begin{figure}
\includegraphics[width=3.5in]{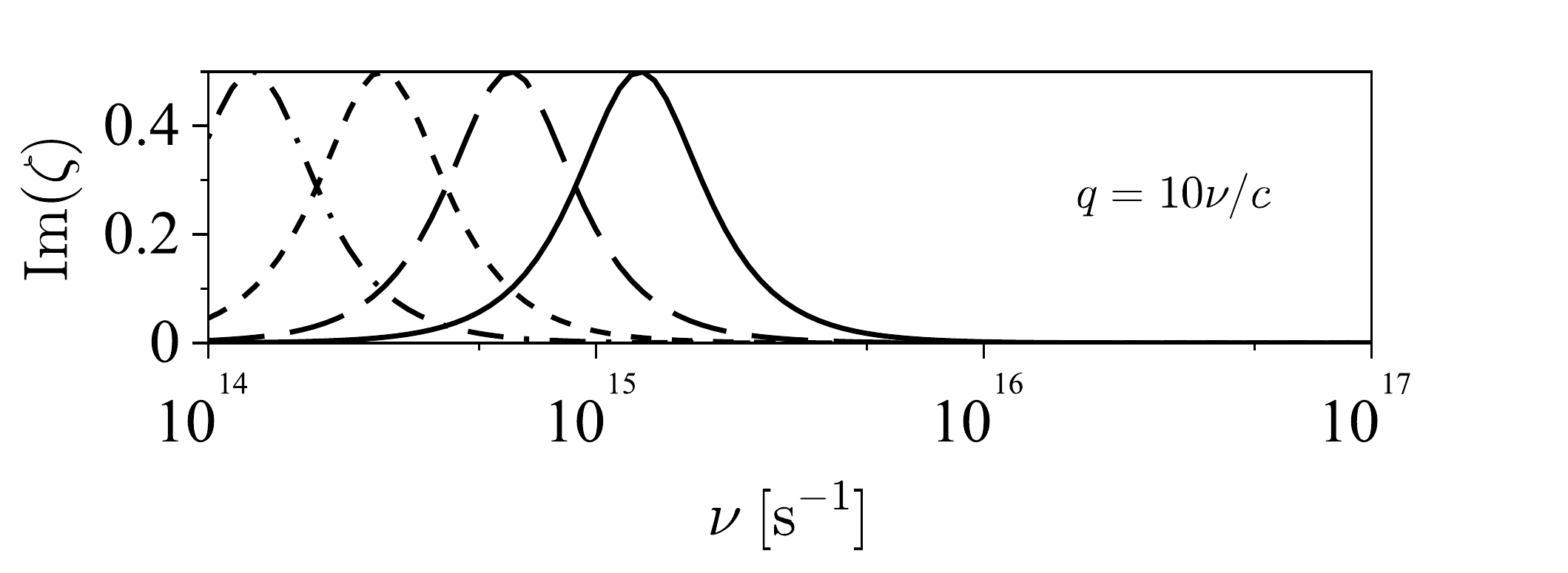}
\caption{The same as Fig. 2 but $q=10\nu/c$. }
\end{figure}
If $\nu$ is fixed, for instance, $\nu=3\times10^{18}$ Hz, then for any $\omega$ between 0 and $10^4$ Hz, the electrons do not screen the proton supercurrent at any length scale shorter than the electron mean free path ($q>\nu/c\approx10^8\;\mathrm{cm}^{-1}$).
A somewhat nontrivial result is that increasing of $\omega$ improves the screening effectiveness; for example, with the collision rate $\nu=10^{17}$ Hz, we would have complete screening ($\zeta=-1$) at $q=\nu/c\approx3\times10^6\;\mathrm{cm}^{-1}$ only for $\omega\geq10^8$ Hz (this numerical result is not shown explicitly in the Figures but could be seen in Fig. 1 as ultimate shift of the curve $\zeta$ to the right leaving in the plot only the ``tail'' with $\zeta=-1$).
One can say that increasing of $\omega$ at fixed $q$ moves the curve $\zeta$ in Fig. 1 to the right.

The imaginary part of $\zeta$ is shown in Figs. 2 and 4.
It is zero at either complete screening ($\zeta=-1$) or at screening absent ($\zeta=0$), while in the intermediate case, which can be called an incomplete screening, the induced current may have the phase-shifted (by $\pi/2$) component with magnitude equal to that of the in-phase component of the induced current.

\emph{Conclusions}.
\begin{figure}
\includegraphics[width=3.1in]{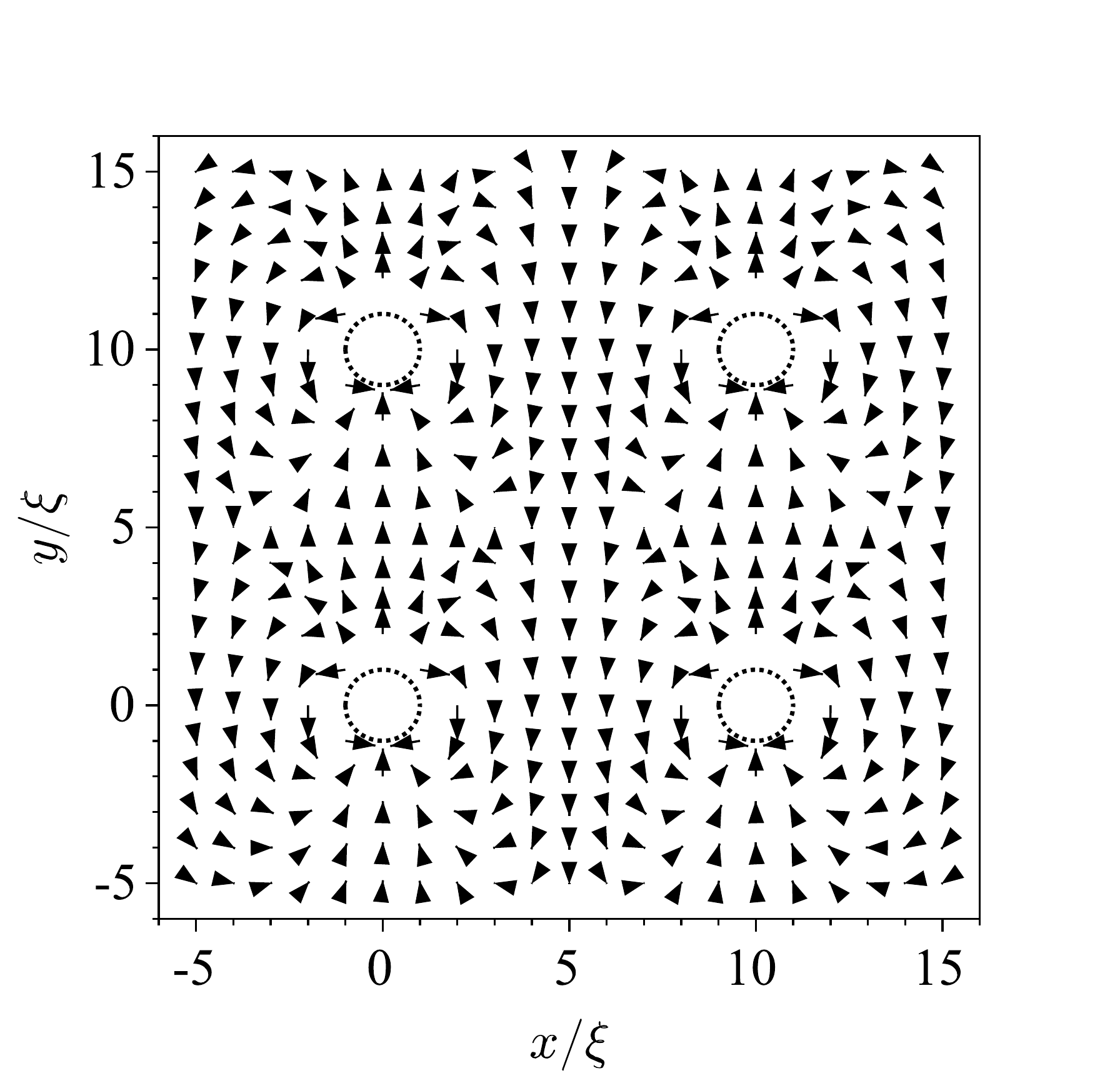}
\caption{Arrows show the vector pattern of the unscreened (without the electron contribution) microscopic electric field $\mathbf{E}_{\mathrm{tube}}(\mathbf{r})$ due to a moving array of four vortices. The field $\mathbf{E}_{\mathrm{tube}}(\mathbf{r})$ is calculated from Eq. (\ref{evortex}) with $\lambda/\xi=10$ corresponding to $\Delta=0.8956$ MeV in Eqs. (\ref{lambda0}) and (\ref{xi0D}) and $d_v=\lambda$ corresponding to $H=3.231\times10^{15}$ Oe in Eq. (\ref{dv}). The normal cores of the flux tubes are displayed by the dotted circles. The velocity of flux tubes relative to the electron liquid is pointed towards $-\mathbf{e_x}$ direction.}
\end{figure}
Based on microscopic physics, I have developed the framework, which can be used to estimate the effectiveness of the electrical current response of the electrons to a given proton supercurrent in presense of a lattice of flux tubes.
I have used typical parameters corresponding to the core of neutron stars and have studied the screening condition for various values of the electron momentum-nonconserving collision frequency.
I found that for typical frequencies of change of the relative velocity between the electrons and the superconducting protons (between about 0 and $10^4$ Hz), the electrons are unable to screen the proton supercurrent if the latter is present.

The presence of the proton supercurrent in the spatial regions between the flux tubes is easy to see.
In realistic conditions, the dynamics arises because the proton flux tubes are forced to move relative to the electrons.
For example, the low-lying sound mode in the outer core corresponds to out-of-phase motion of the neutron fluid and the electron-proton plasma \cite{KobyakovEtAl2017}, while the total matter density is locally unperturbed.
In case when the proton flux tubes and the neutron vortices are present, this mode will be modified, but will be still possible.
Due to the vortex-flux tube pinning, in that mode one expects that the flux tubes are guided by the motion of the neutron vortices.
If the velocity of this mode is of the order of $0.1c$, then its lowest frequency (when its wavelength is of the order of the stellar size $10^6$ cm) is of the order of 3 kHz, which is in the range of frequencies mentioned above and considered in Figs. 1-4.

It is easy to see, using the first London equation \cite{Tinkham}, why a motion of the flux tube relative to the electrons induces the microscopic electric field (which in turn might induce the electric current depending on the electric conductivity).
The electric supercurrent of a single proton flux tube is given by
\begin{equation}\label{Jtube}
  \mathbf{J}_{\mathrm{tube}}(\mathbf{r})\approx Y_0\left(\mathbf{e_y}\frac{x}{r^2}-\mathbf{e_x}\frac{y}{r^2}\right)e^{-r/\lambda},
\end{equation}
where $Y_0=n_{p}{\hbar e}/{2m_p}$, $(\mathbf{e_x},\mathbf{e_y},\mathbf{e_z})$ are Cartesian basis vectors, $\mathbf{e_z}$ is along the flux tube, $\mathbf{r}=\mathbf{e_x}x+\mathbf{e_y}y$ and $r^2=x^2+y^2$.
In the rest frame of reference of the electrons, a moving flux tube with velocity $\mathbf{v}_L=\mathbf{e_x}v_L$ generates the microscopic electric field, which can be easily found from the first London equation \cite{Tinkham}:
\begin{equation}\label{evortex}
  \mathbf{E}_{\mathrm{tube}}(\mathbf{r})=\frac{m_p}{n_pe^2}\partial_t\mathbf{J}_{\mathrm{tube}}(\mathbf{r}-\mathbf{v}_Lt)=-(\mathbf{v}_L\cdot\nabla')\mathbf{J}_{\mathrm{tube}}(\mathbf{r}'),
\end{equation}
where $\nabla'=\partial_{\mathbf{r}'}$ and ${\mathbf{r}'=\mathbf{r}-\mathbf{v}_Lt}$.
Inserting Eq. (\ref{Jtube}) into (\ref{evortex}) one finds $\mathbf{E}_{\mathrm{tube}}(\mathbf{r})$.

In Fig. 5, I plot the microscopic electric field $\mathbf{E}_{\mathrm{tube}}(\mathbf{r})$ due to a moving array of four vortices at a fixed instant of time.
One observes that the transverse electric field is indeed generated in the spatial region between the vortices.
In case when the system parameters lead to $\zeta=0$ (this is the case for the realistic parameters as discussed in detail above), the vortex-induced electric field generates the proton electric current in the spatial region between the vortices, while the electron current is not excited; in this case the screening condition is not satisfied.
In the opposite case when $\zeta=-1$, the electron current is excited and completely screens the proton supercurrent, so the resulting total electric current is zero; however this case is not realized for the realistic parameters.
The astrophysical implication is that since the screening condition is not satisfied, then the earlier conclusions regarding the electron-flux tube interactions in neutron star core must be reassessed; the rate of momentum exchange between the electrons and the flux tube lattice in the superconducting and/or superfluid nuclear matter in the core of neutron stars remains an open question.

\emph{Acknowledgements.} This work was supported by the Center of Excellence ``Center of Photonics'' funded by The Ministry of Science and Higher Education of the Russian Federation, contract No. 075-15-2022-316.

\end{document}